%% file: main.tex
\documentclass[sigconf, authorversion, nonacm]{acmart}

\usepackage{multirow}

\AtBeginDocument{%
  \providecommand\BibTeX{{%
    \normalfont B\kern-0.5em{\scshape i\kern-0.25em b}\kern-0.8em\TeX}}}

\setcopyright{acmcopyright}
\copyrightyear{2018}
\acmYear{2018}
\acmDOI{10.1145/1122445.1122456}

\acmConference[CHI '21]{CHI '21: ACM conference on Human Factors in Computing Systems}{May 08--13, 2021}{Yokohama, Japan}

\begin{document}

\input{Sections/Title.tex}
\renewcommand{\shortauthors}{Suzuki et al.}

\input{Sections/Abstract.tex}



\begin{teaserfigure}
  \centering
  \includegraphics[width=\textwidth]{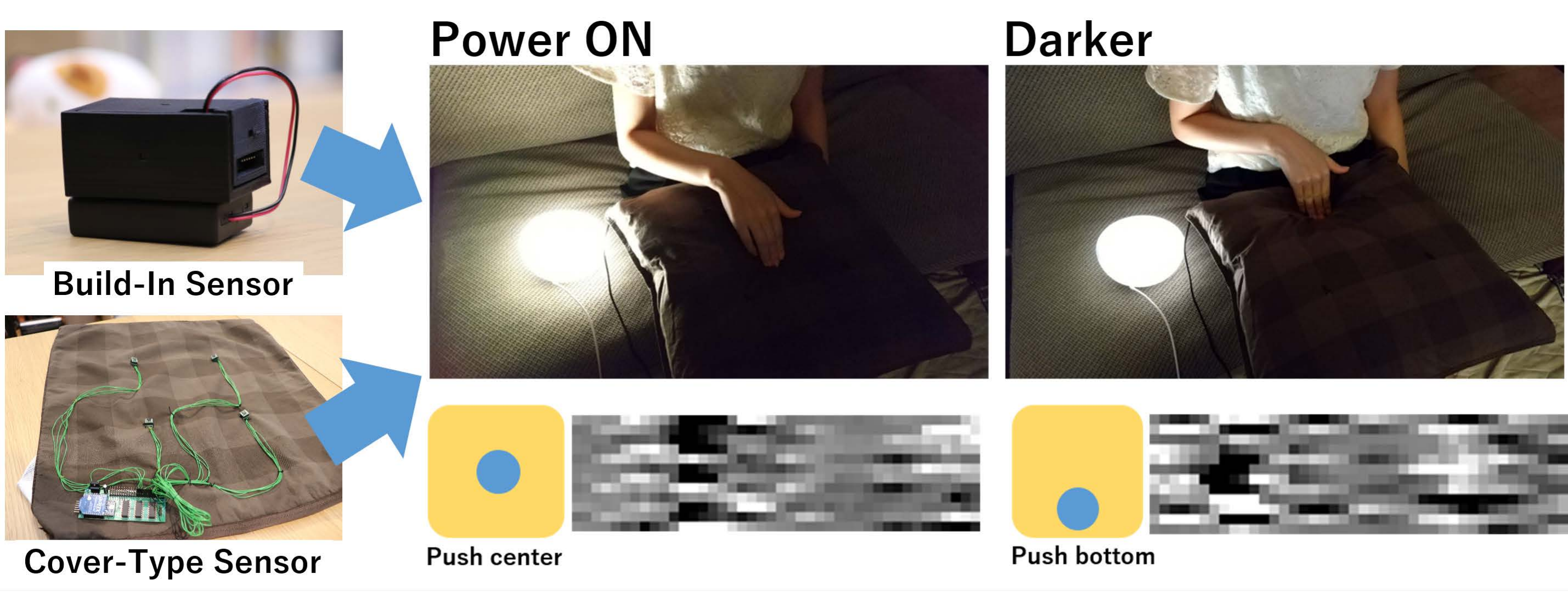}
  \caption{Overview of Cushion Interface}
  \label{fig:fig1}
\end{teaserfigure}

\maketitle

\input{Sections/Intro}

\input{Sections/RelatedWork}

\input{Sections/Design}

\input{Sections/Estimation}

\input{Sections/Evaluation}

\input{Sections/Discussion}

\input{Sections/Conclusion}


\bibliographystyle{ACM-Reference-Format}
\bibliography{reference.bib}

\appendix








\end{document}

%% file: Sections/Title.tex
\title[Smart-home Control by Cushion Interface]{Exploring Gestural Interaction with a Cushion Interface for Smart Home Control}


\author{Yuri Suzuki}
\email{s.yuri23118@keio.jp}
\affiliation{
  \institution{Keio University}
  \city{Yokohama}
  \country{Japan}
}

\author{Kaho Kato}
\email{kaho\_0128@keio.jp}
\affiliation{
    \institution{Keio University}
    \city{Yokohama}
    \country{Japan}
}

\author{Naomi Furui}
\email{akonako\_1012@keio.jp}
\affiliation{%
  \institution{Keio University}
  \city{Yokohama}
  \country{Japan}
}

\author{Daisuke Sakamoto}
\email{sakamoto@ist.hokudai.ac.jp}
\affiliation{%
 \institution{Hokkaido University}
 \city{Yokohama}
 \country{Japan}
}

\author{Yuta Sugiura}
\email{sugiura@keio.jp}
\affiliation{%
    \institution{Keio University}
    \city{Yokohama}
    \country{Japan}
}




%% file: Sections/Abstract.tex
\begin{abstract}

In this research, we aim to realize cushion interface for operating smart home. We designed user-defined gestures using cushion and developed gesture recognition system. We asked some users to make gestures using cushions for operating home appliances and determined user-defined gesture sets. We developed two methods for gesture identification. The First, We inserted sensor modules consisting of photo reflective sensors and acceleration sensor inside a cushion. The second, we embedded the acceleration sensor arrays in the cushion cover. Gesture recognizer was implemented using Convolutional Neural Networks (CNN). To evaluate our method, We conducted an experiment to measure recognition accuracy.
Results showed that an average accuracy was 94.8\% when training for each user, and an average accuracy of 91.3\% when testing with a user that did not exist in the training data set. 
\end{abstract}

%% file: Sections/Intro.tex
\section{introduction}
There is an abundance of research on the ubiquitous computing that is embedded in all kinds of familiar objects, using computers to support people’s daily lives~\cite{Kidd, Brumitt}. Currently, smart homes and other connected information systems (e.g., Internet of Things [IoT]) are becoming popular, with computers built into home appliances. These information appliances are operated using various types of user interfaces. We can expect future information appliances to be operated in various ways as well, and with each new device, users will need to learn these operations. In other words, as the number of such appliances in a home increases, the number of remote controls and other interfaces will increase. It will take time to learn which interface is for which device and how to use them, and this will place an increasing burden on users. For this reason, there is a desire for an integrated interface that can be used to operate a diverse range of appliances.

There has been earlier research on interfaces for integrated operation of information appliances. For example, Tsukada et al. developed Ubi-Finger as a method of operating devices using gestures, controlling multiple appliances by attaching devices to the fingertips~\cite{Tsukada}. Ubi-Finger enables intuitive control by using gestures that are similar to real-world operation of the device, but the constant need to have devices attached to the fingers in normal living situations could impose a psychological burden on users. Intelligent Room~\cite{Irie} uses multiple cameras to convert an entire room into an intelligent robot and is able to operate appliances based on gestures. Users make everyday natural motions without having to wear anything, but introducing cameras into people’s living spaces may involve issues such as screening and infringement of privacy. Another approach is to use a familiar everyday object~\cite{Masui, Seki}. Natural interfaces that easily become familiar can be designed by using objects that are already in our living spaces and that have a known identity~\cite{Shiio}. A tabletop touch interaction display is another alternative for future home control~\cite{10.1145/1731903.1731911}.

Among the huge design space for new home control devices, we have focused on cushions, which are a common and familiar everyday item often found in living rooms and bedrooms, in particular. People interact with cushions in various ways every day, hugging them, squeezing them, hitting them, and more. Our idea is to use these interactions as gestures to build interfaces for home appliances, with a soft touch that would blend easily into our lives.
Although we present a new remote for controlling appliances, the cushion interface is envisioned as something to be used together with existing input methods rather than something to replace them. This is because existing interfaces may be suited to various conditions in people’s living spaces, as factors like location and posture change dynamically. Cushions also have the benefit of being within reach from the sofa and other areas intended for relaxation, so operations can be completed  while relaxing. Such interfaces do involve some cost, but the barrier to introducing them should be low, because the interface functionality will come without losing any original functionality as an everyday item in the living environment. These interfaces would not attempt to support all operations of an information device, but rather, they could be practical in everyday use by implementing the most frequently occurring operations.


In this study, we designed a set of gestures optimized for using a cushion to control household appliances and developed a system for recognizing time sequence data of these gestures. Our controller operates appliances such as televisions and lighting by performing actions such as pushing or turning over a cushion. We first conducted a survey to ask several users for suggestions of  cushion gestures that would correspond to operations on each appliance, and from these, a user-defined gesture set was selected. We then prototyped two types of sensing systems. The fist sensor type was built by embedding a sensor module composed of reflective optical sensors and acceleration sensors in a cushion, with reference to an existing method~\cite{Sugiura}. The second sensor type is a sensor module that is attached to the cushion cover. We implemented gesture recognition using a convolutional neural network (CNN) and then evaluated the accuracy of recognition. The contributions of this research are as follows.
\begin{itemize}
    \item A gesture input method has been designed for a square cushion that operates appliances found in an everyday living environment through a user-defined gesture approach~\cite{Jacob}.
    \item A system was developed using a CNN to recognize a user-defined gesture set, employing sensors embedded in a cushion or attached to a cushion cover.
\end{itemize}

%% file: Sections/RelatedWork.tex
\section{Related work}

\subsection{Home Appliance Interfaces}
In the field of human–computer interaction, research on gestural interaction has been actively conducted examining how to operate home appliances~\cite{Yoshizawa, Ujima, Gonzalo}. Starner et al. A paper by Starner et al. proposed the Gesture Pendant~\cite{Starner},  which is a gesture-recognition device encompassed in a pendant, using infrared LEDs and a camera. The concept of the system is a user wears the pendant around their neck and performs gestures in front of it to control lighting, a television, or other home appliances. Similarly, with Ubi-Finger~\cite{Tsukada}, the user wears a device on a finger, and the device recognizes gestures that are used to control home appliances. These gestures incorporate motions similar to real-world operation of the devices, such as turning the volume knob on an audio device or pushing the power button on a television, providing intuitive control of appliances. Those interfaces work well to control appliances at home, but users of those interfaces must wear a device at all times to control their appliances. Conversely, we are interested in not wearing devices; the key idea of our work is to use a cushion, which is a common item in homes, as a touch interaction device. We also focused on designing gestural interactions to control home appliances because they would be a natural way to operate appliances in daily life. One thing that is important to mention is that we do not aim to replace dedicated remotes for controlling appliances. Our interface is a new alternative for controlling appliances and will play a supplementary role in daily activities by allowing people to control common uses of an appliances with a device that is readily at hand.

\subsection{Soft Object Interfaces}

Our lives are filled with soft objects. Sofas and chairs, carpets, curtains, beds, pillows, stuffed toys, and cushions are many of the objects frequently nearby people that are soft objects. In prior studies of soft objects as an interface for computing devices, Sugiura et al. incorporated multiple optical sensor modules into deformable everyday objects such as cushions and used center-of-gravity calculations to estimate points of contact and machine learning to estimate the user’s posture~\cite{Sugiura}. They also developed a cushion interface by dividing the surface of the cushion into several sections and operating individual home media devices by tapping in each of these sections.
Ikeda et al. proposed a cushion interface called FunCushion with push input and output functionality~\cite{Ikeda}. They referenced the Sugiura et al. method of input detection using infrared light, and added a light-emitting function to the cushion using fluorescent materials to present information on the surface of the cushion. In the same vein, Media Cushion incorporates touch sensors on the outside and optical sensors on the inside of the cushion, allowing the cushion to be used to make changes in the living environment, such as to lighting or audio, in a natural way~\cite{Yagi}. The system was developed based on the idea of inferring the state and intentions of the user from unconscious behaviors and providing an appropriate environment, so it does not require the user to intentionally control appliances. 

Soft household items other than cushions, such as stuffed toys,  have also been used to develop interfaces~\cite{Johnson, Aabeele}. Paiva et al. developed a method of controlling the emotional states of a synthetic character in a computer game with a tangible interface named SenToy, which is a doll with sensors ~\cite{Paiva}. The prototype material can be incorporated into items such as stuffed toys, to create a remote control that can operate appliances such as lighting or an air conditioner in a relaxed environment. Seki et al. also developed a stuffed toy interface for smart homes, to be used in areas intended for relaxing, such as living rooms~\cite{Seki}. The interface uses accelerometers and reflective optical sensors to control a television through actions such as shaking or grasping the limbs of the stuffed toy. Vanderloock et al. developed a soft interface filled with conductive fibers~\cite{Vanderloock}. The item contains many electrodes, and the system learns changes in the conductivity between these electrodes when pressure is applied, deforming the item. This enables the system to recognize gestures such as ``curl'' or ``stretch.'' 

Those studies demonstrate home appliance control with new technology such as sensors and devices that use a wide variety of interactions. In the present study, we focused on the cushion as a soft object for smart home control. The key idea is to take the user-defined gesture approach in a kind of participatory design process; this is the major contribution of this study as compared to the work by Sugiura et al. ~\cite{Sugiura}. We created the new cushion-based remote and interface based on user feedback, which makes the interface acceptable to the people who will use the interface at home.

\subsection{User-Defined Gesture Approach}

User-defined gestures refers to asking users to come up with mannerisms or movements that are suitable for each operation and then designing a device’s input gestures based on their input~\cite{Jacob}. The users do not have preconceptions about how to use the device, so it is easy to obtain gestures that are natural. This approach has been used for design in earlier research, including in a system for large, foot-based gestures~\cite{Alexander} used in Fukahori et al.’s plantar-based foot gesture system for operating mobile computer terminals~\cite{Fukahori} and with a smart home gesture system using smartphones~\cite{Kuhnel}. In this study, we also relied on user-defined gestures by having participants propose suitable gestures to use with a cushion for controlling specific devices.


\subsection{Gesture Recognition by Visualizing Time-Sequence Data}

We developed a gesture-recognition technique for understanding movements that users will make on the cushion. Machine learning techniques are used for this purpose in general, and in particular, support vector machines (SVM) are a machine learning technique often used in prior human-computer interaction (HCI) research. SVMs have been widely used because they are easy to use, but fast Fourier transform (FFT) preprocessing is generally necessary to recognize time sequence data, so FFT results are often used as feature values. Our research uses 20 reflective optical sensors and a three-axis accelerometer, requiring dozens of feature values for each sensor and adding complexity to data preprocessing. Recent gesture-recognition research has proposed methods that visualize sensor data, lining up time sequence sensor data as a gray-scale image in order to learn feature values~\cite{Fukui, Kikui}. Fukui et al. developed a device able to recognize hand gestures by measuring wrist indentations using sensors attached to the arm~\cite{Fukui}. Their study created images using data from range sensors arranged inside the device, and then uses an SVM to recognize gestures from histogram of oriented gradients (HOG) feature values. Kikui et al. converted time sequence sensor data obtained from a glasses-type device with embedded optical sensors into 2D rectangular data and used it to train a CNN to detect gestures of the face (cheeks) and to reduce the cost of user training~\cite{Kikui}. Our research also converts time sequence sensor data to images and recognizes gestures using a CNN trained with such data.

%% file: Sections/Design.tex
\section{Gesture Elicitation Study}
We designed a gesture set to make a cushion an interface for controlling appliances. However, the kind of gesture set that will be appropriate for users in their everyday lives at home is not obvious. So we took a user-defined gesture approach to a kind of participatory process involving users in the design steps. Similarly, what types of appliances are commonly in users’ homes is also not obvious. So we conducted a survey to understand which appliances might be controlled by the cushion interface. We designed a gesture set based on the first survey and on a second user-involved design process.


\subsection{Survey for Selecting Appliances and Operations for Designing Gestures}

We first conducted a survey to decide what home appliances to work with in our research. We surveyed 20 participants (11 female) aged from 21 to 59 (average age, 26) regarding appliances often found in a living room. The results included 20 votes for television, 20 for lighting, 19 for air conditioner, 11 for audio devices, and 9 for a fan, and a majority of the participants possessed these appliances. Accordingly, we selected those five devices (air conditioner, audio device, television, lighting, and fan) for creating gesture operations.

We then surveyed 10 participants (5 female) aged 21 to 25 (average age, 23), including participants from the previous survey, for common operations of these five appliances. This yielded a total of 48 operations. The 24 operations that were indicated by 30\% or more of participants are shown in Table \ref{tab:table1}. Note that action pairs are separated into two operations (e.g. ``adjust temperature'' is divided into ``raise temperature'' and ``lower temperature''), while cases where the same action performs multiple operations (e.g. ``power ON'' and ``power OFF'' or ``play'' and ``stop'') are consolidated into one operation.

\begin{table}[htb]
  \centering
  \caption{Results of questionnaire on home appliance operation}
  \begin{tabular}{|l|l|}
  \hline
Appliance                       & Operation              \\ \hline
\multirow{5}{*}{Air conditioner} & Power ON/OFF       \\ \cline{2-2} 
                                & Switch mode        \\ \cline{2-2} 
                                & Raise temp.        \\ \cline{2-2} 
                                & Lower temp.        \\ \cline{2-2} 
                                & Set timer          \\ \hline
\multirow{6}{*}{Audio device}   & Power ON/OFF       \\ \cline{2-2} 
                                & Play/Stop          \\ \cline{2-2} 
                                & Raise volume       \\ \cline{2-2} 
                                & Lower volume       \\ \cline{2-2} 
                                & Next track         \\ \cline{2-2} 
                                & Previous track         \\ \hline
\multirow{6}{*}{Television}     & Power ON/OFF       \\ \cline{2-2} 
                                & Raise volume       \\ \cline{2-2} 
                                & Lower volume       \\ \cline{2-2} 
                                & Next channel     \\ \cline{2-2} 
                                & Previous channel   \\ \cline{2-2} 
                                & Program guide      \\ \hline
\multirow{3}{*}{Lighting}       & Power ON/OFF       \\ \cline{2-2} 
                                & Brighter           \\ \cline{2-2} 
                                & Darker             \\ \hline
\multirow{4}{*}{Fan}            & Power ON/OFF       \\ \cline{2-2} 
                                & Change speed       \\ \cline{2-2} 
                                & Oscillation ON/OFF \\ \cline{2-2} 
                                & Set timer          \\ \hline
\end{tabular}
  \label{tab:table1}
\end{table}

\subsection{Gesture Elicitation}

In addition to the 24 operations obtained from the survey, we added an operation to switch the device being operated. This was because there are generally multiple devices in a living room, and it will be necessary to decide which appliance to use before performing an operation. For the total of 25 operations, we conducted an elicitation study in which users devised gestures suitable for each operation (Table~\ref{tab:table1}). We had 15 participants (6 female) aged from 21 to 59 (average age, 27), and 11 of these were graduate students in information-related disciplines. All participants were right-handed. Participants were shown the 25 operations one at a time and instructed to think of suitable gestures. Operations were given in random order for each appliance.

\begin{figure}[htb]
  \centering
  \includegraphics[width=0.8\linewidth]{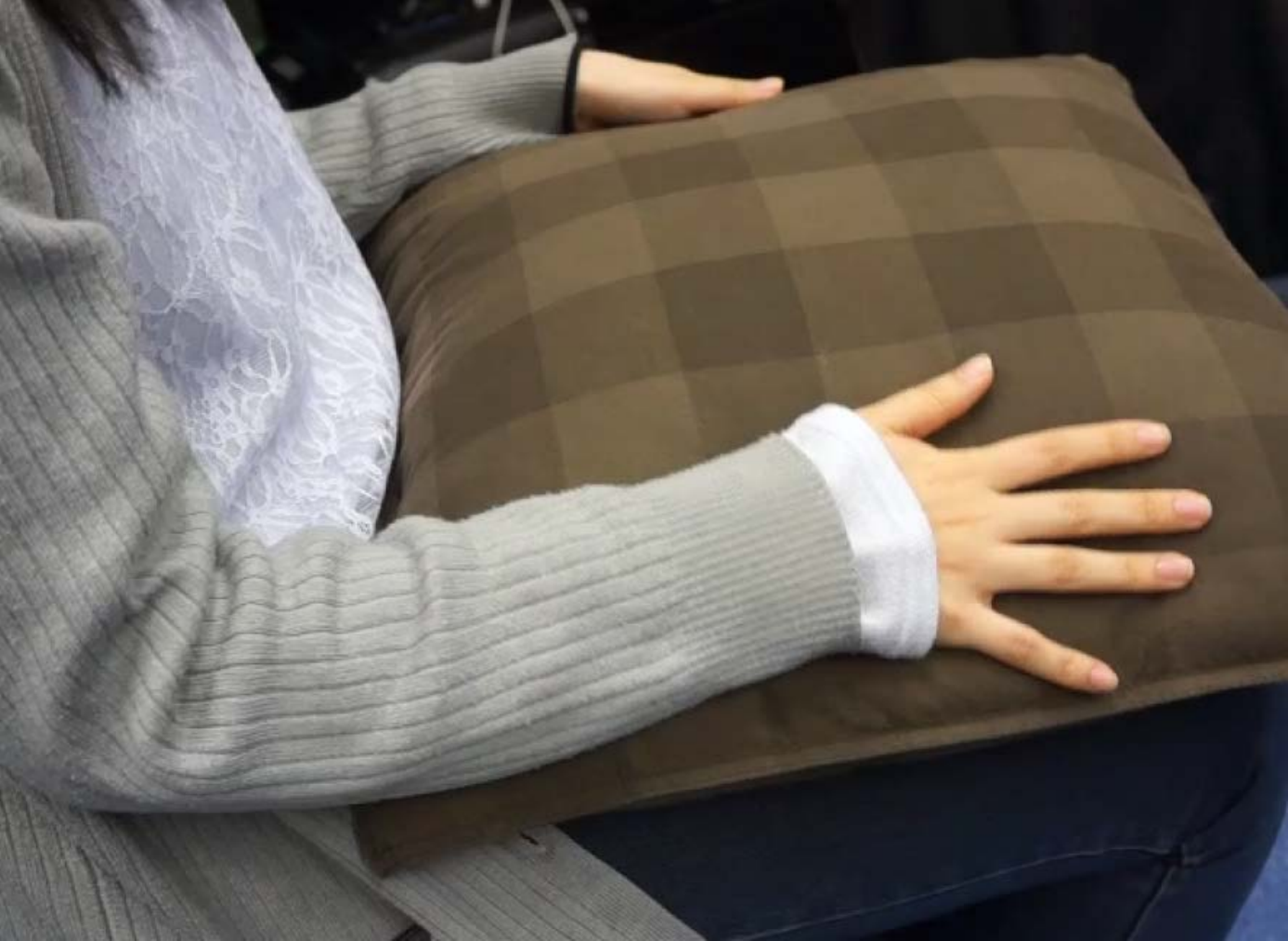}
  \caption{Cushion experiment}
  \label{tab:experiment}
\end{figure}

In the study, participants were given a 43 cm $\times$ 43 cm cushion (Figure \ref{tab:experiment}), and after being shown videos of performing each operation on a PC screen, they were asked to think of an appropriate gesture for the operation using the cushion and to perform the gesture. The study was recorded using a video camera. Note that the cushions used were ordinary, off-the-shelf cushions with no sensors.

Participants were also instructed that they need not consider conflicts among gestures and they need not consider whether gestures would be recognizable. The experiment yielded 15 proposed gestures for each of the 25 operations, totaling 375 possible gestures. Excluding duplicated gestures yielded a total of 98 gestures.

\subsection{Selecting Gestures from Collected Ideas}

We selected a set of cushion gestures from the proposed gestures obtained in the gesture elicitation study, according to the following criteria that the gesture be one that the most participants proposed for each appliance.

As already noted, we assigned gestures with linear symmetry to operation pairs (e.g., ``increase temperature'' and ``decrease temperature''). If there was a conflict among gestures for a particular appliance, we gave priority to the gesture proposed by the most participants and assigned the gesture that was in second place to the other operation.

\begin{figure}[htb]
  \centering
  \includegraphics[width=\linewidth]{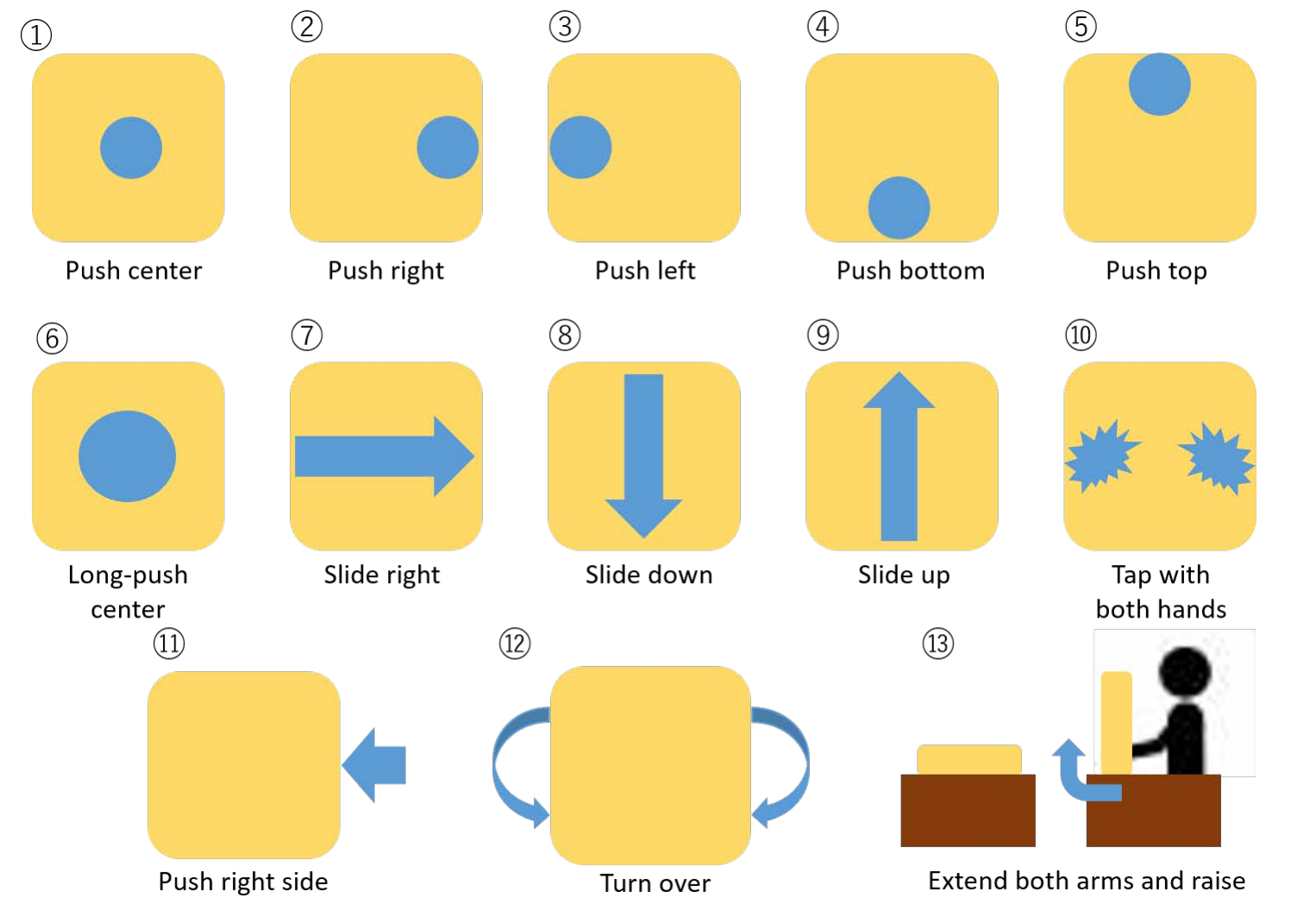}
  \caption{Types of gestures}
  \label{tab:gesture}
\end{figure}

The gesture set obtained is shown in Table~\ref{tab:table2}. Excluding duplications, there were a total of 13 gestures  (Figure~\ref{tab:gesture}). Several gestures required that the cushion be divided into sections and involved pushing or tapping one of the sections. We presume that such gestures tended to be suggested because participants often perform operations by pushing buttons on devices such as remote controls or by tapping on screens. There were also participants who suggested gestures on the cushion similar to performing operations on real-world objects, such as a smartphone or newspaper. On the other hand, the ``turn over'' gesture is unusual among user interfaces (UIs) and could be considered particular to cushions. Pairs of operations such as ``slide up'' and ``slide down'' tended to be suggested for operation pairs like increasing and decreasing the volume or temperature. ``Slide right'' was the most frequently proposed operation for switching modes on the air conditioner. There was no operation to pair with this, so ``slide left'' was not assigned. In general, ``slide left'' was proposed less often than ``slide right,'' which we suppose may be because the participants were right-handed, which could make the slide left gesture more difficult to perform.

\begin{table*}[htb]
  \centering
  \caption{Result of the gesture elicitation study}
\begin{tabular}{|l|l|l|r|}
\hline
Appliance                        & Operation          & Gesture                    & \begin{tabular}[c]{@{}l@{}}No. of \\ participants \\ proposing\end{tabular} \\ \hline
\multirow{5}{*}{Air conditioner} & Power ON/OFF       & Push center                & 7                             \\ \cline{2-4} 
                                 & Switch mode        & Slide right                & 3                             \\ \cline{2-4} 
                                 & Raise temp.        & Slide up                   & 4                             \\ \cline{2-4} 
                                 & Lower temp.        & Slide down                 & 4                             \\ \cline{2-4} 
                                 & Set timer          & Push right side            & 2                             \\ \hline
\multirow{6}{*}{Audio device}    & Power ON/OFF       & Long-push center           & 4                             \\ \cline{2-4} 
                                 & Play/Stop          & Push center                & 8                             \\ \cline{2-4} 
                                 & Raise volume       & Push top                   & 3                             \\ \cline{2-4} 
                                 & Lower volume       & Push bottom                & 3                             \\ \cline{2-4} 
                                 & Next track         & Push right                 & 5                             \\ \cline{2-4} 
                                 & Previous track     & Push left                  & 5                             \\ \hline
\multirow{6}{*}{Television}      & Power ON/OFF       & Push center                & 7                             \\ \cline{2-4} 
                                 & Raise volume       & Push top                   & 5                             \\ \cline{2-4} 
                                 & Lower volume       & Push bottom                & 5                             \\ \cline{2-4} 
                                 & Next channel       & Push right                 & 6                             \\ \cline{2-4} 
                                 & Previous channel   & Push left                  & 6                             \\ \cline{2-4} 
                                 & Program guide      & Extend both arms and raise & 2                             \\ \hline
\multirow{3}{*}{Lighting}        & Power ON/OFF       & Push center                & 8                             \\ \cline{2-4} 
                                 & Brighter           & Push top                   & 5                             \\ \cline{2-4} 
                                 & Darker             & Push bottom                & 5                             \\ \hline
\multirow{4}{*}{Fan}             & Power ON/OFF       & Push center                & 9                             \\ \cline{2-4} 
                                 & Change speed       & Push top                   & 4                             \\ \cline{2-4} 
                                 & Oscillation ON/OFF & Tap with both hands        & 2                             \\ \cline{2-4} 
                                 & Set timer          & Push right                 & 2                             \\ \hline
All appliances                   & Switch device      & Turn over                  & 4                             \\ \hline
\end{tabular}
  \label{tab:table2}
\end{table*}

%% file: Sections/Estimation.tex
\section{Gesture Recognition}

Next, we turned a cushion into an interface device with sensors and machine learning techniques and developed two sensors for receiving user input in terms of how the cushion would be touched and stroked. For the gesture recognition, we employed a CNN-based machine learning technique. We first gathered time sequence data from sensors in grayscale images for use with the image-based gesture-recognition task in the machine learning technique.

\subsection{Hardware}

We create two types of devices for gesture recognition on the cushion. One was a built-in sensor, and the other was a cover-type sensor. The built-in sensor is a unit that will be embedded in the cushion as in prior studies~\cite{Sugiura}. It employs four outward-facing reflective optical sensors to capture low data of how a user touches and strokes the cushion. The second type of device, the cover-type sensor, is designed to solve the shortcomings of the first sensor being non-removable from the cushion; the built-in sensor was placed by cutting the cushion open and embedding the sensor unit into the inner foam of the cushion, which could be disappointing to people who may be unhappy at having to cut their cushion open in the even repairs or replacement is needed. Thus, for the cover-type sensor, we sewed four acceleration sensors onto the inner surface of a cushion cover, making it easy to install sensors as compared to the built-in sensor.


\subsubsection{Build-in sensor}

For the built-in sensor, we created sensor modules to be incorporated into cushions, referring to earlier methods~\cite{Sugiura} (Figure 4, upper left). We created an outer case (44.3 mm $\times$ 68.3 mm $\times$ 33.0 mm) using a 3D printer, and attached an outward-facing reflective optical sensor (Kodenshi Corp. SG-105, Kodenshi Corp., Kyoto, Japan) to the center of each of four sides, excluding the bottom. The case housed a microcontroller (Arduino Pro Mini, Arduino, Somerville, MA, USA) and a wireless communication device (Xbee, Digi, Hopkins, MN, USA) in anticipation of future independent operation. A battery was attached outside the case.

\begin{figure}[htb]
  \centering
  \includegraphics[width=0.8\linewidth]{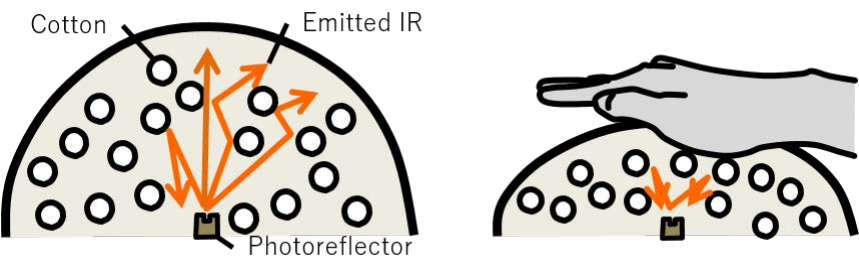}
  \caption{Principle of the build-in sensor \cite{Sugiura}}
  \label{tab:buildin}
\end{figure}

\begin{figure}[htb]
  \centering
  \includegraphics[width=\linewidth]{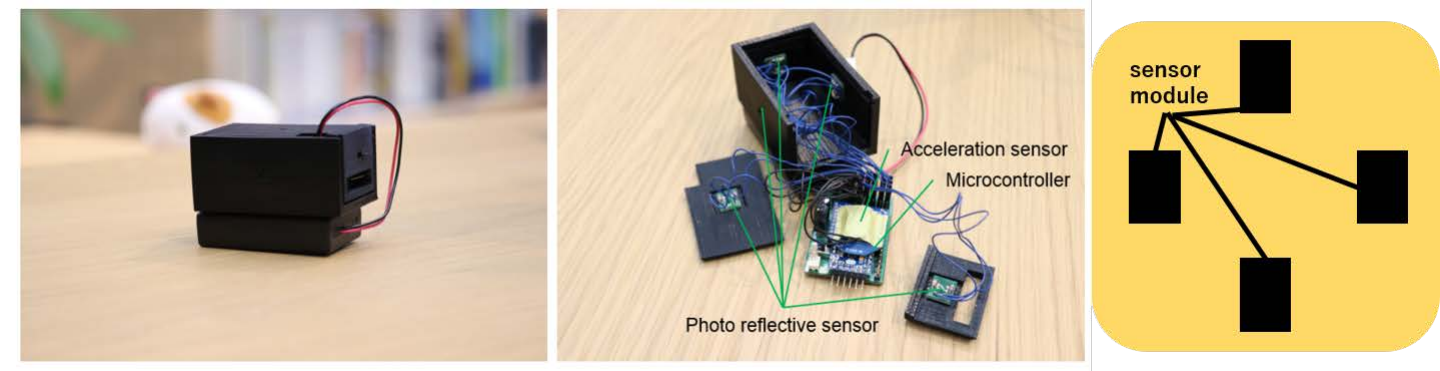}
  \caption{Sensor module (left), components (middle), and sensor module arrangement (right)}
  \label{tab:module}
\end{figure}

The reflective optical sensors are a monolithic device with an infrared LED as an emitter and a phototransister as a receiver. The amount of emitted infrared light that is reflected back is converted to a sensor value and can be used to compute the density of the cotton material when the sensor is placed in a cushion. Applying pressure to the cushion changes the local density of cotton inside the cushion (Figure~\ref{tab:buildin}), so that gestures such as ``push'' or ``tap'' can be detected. To detect changes in the cushion such as ``turn over'' or ``hold up,'' an acceleration sensor was also included in one of the four modules (Figure~\ref{tab:module}, upper right). The microcontroller uses the accelerometer to read values for the x, y, and z directions. As a result, a single cushion produces sensor data for 23 dimensions (5 $\times$ 4 = 20 for reflective optical sensors and three for the accelerometer; see Figure~\ref{tab:module}). 

\subsubsection{Cover-Type Sensor}

\begin{figure}[htb]
  \centering
  \includegraphics[width=0.8\linewidth]{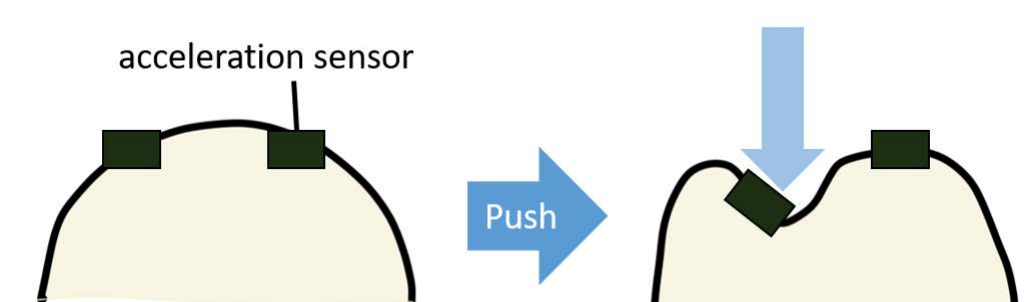}
  \caption{Principle of the cover-type sensor}
  \label{tab:covertype}
\end{figure}

\begin{figure}
  \centering
  \includegraphics[width=\linewidth]{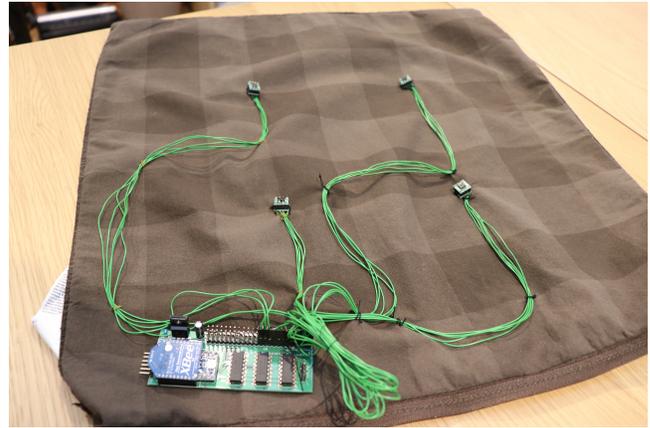}
  \caption{Cushion cover sensor}
  \label{tab:cover}
\end{figure}

For this method, we sewed four three-axis acceleration sensors onto the inner surface of a cushion cover and put the cover on the existing cushion (Figure~\ref{tab:cover}). The microcontroller (Arduino Pro Mini) was connected to each acceleration sensor by a cable and then stored in the cushion. An acceleration sensor can get three values for the x, y, and z directions. Thus, a single cushion produces sensor data for 12 dimensions. When the surface of the cushion is deformed, the value of the sensor changes (Figure~\ref{tab:covertype}). Xbee was deployed in anticipation of using wireless in the future, but in this study, we connected the sensors to a computer with a 1 m USB cable.

\subsection{Collecting Training Data}
We used the CNN to identify the gestures as described in a related study~\cite{Kikui}. In order to use the CNN, the obtained sensor values were converted to grayscale images. Because of the different types of sensors used, there are different ways to handle the sensor values.

\subsubsection{Collecting Training Data with Build-In Sensor}

A calibration process was required for both sensors to set the range of values obtained from each. Users were requested to push, tap, and turn over the cushion to obtain maximum and minimum values. These were then normalized to a maximum value of 1 and a minimum value of 0, which were used as the output values for each sensor. Then, differences between the current output value and those from the previous 10 frames for each sensor were arranged in a time sequence and converted to a greyscale image.


The start of a gesture was taken to be the point where the difference between the output values and those of the prior five frames exceeded a certain threshold. The difference values for 60 frames following the start of a gesture was taken as one gesture. Note that the sampling rate of the sensor module was approximately 20 Hz, so approximately three seconds of data were collected for a single gesture.

\subsubsection{Collecting Training Data with Cover-Type Sensor}
Differences between the current output values and the values from the previous frames for each sensor were arranged in a time sequence. After applying the obtained output values for the low pass filter, min–max normalization was performed so that the maximum value was 255 and the minimum value was 0, thereby converting the obtained output values into a greyscale image. The difference in values of 42 frames (1.2 seconds) following the start of a gesture were taken as one gesture.

\begin{figure}
  \centering
  \includegraphics[width=\linewidth]{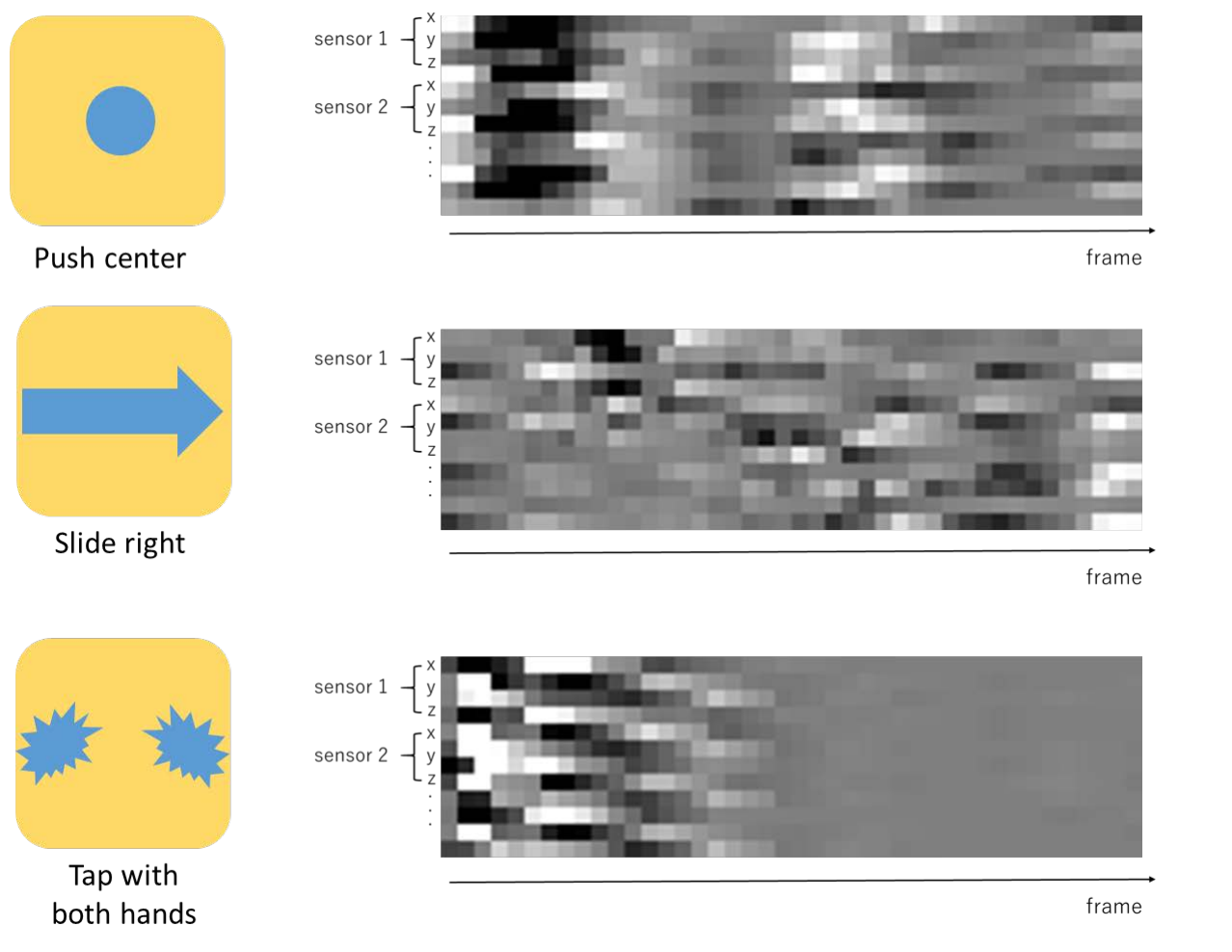}
  \caption{Grayscale images from difference values for three gestures (cover-type sensor).}
  \label{tab:sensor}
\end{figure}

Comparing the grayscale images of gestures, we can see that the patterns in the images differ for each gesture. As examples, the images created by performing the gestures ``push center,'' ``long push center,'' and ``slide right'' are shown in Figure~\ref{tab:sensor}. The vertical axis of each image shows the output values of all sensors, the horizontal axis is the frame number, and the pixel values correspond to difference values. These patterns can be expected to change with the location, force, and speed of a push, as well as with the orientation of the cushion. We attempted to recognize the gestures by using machine learning to classify such images for each gesture.

\subsection{CNN Training}

\begin{figure}
  \centering
  \includegraphics[width=0.4\linewidth]{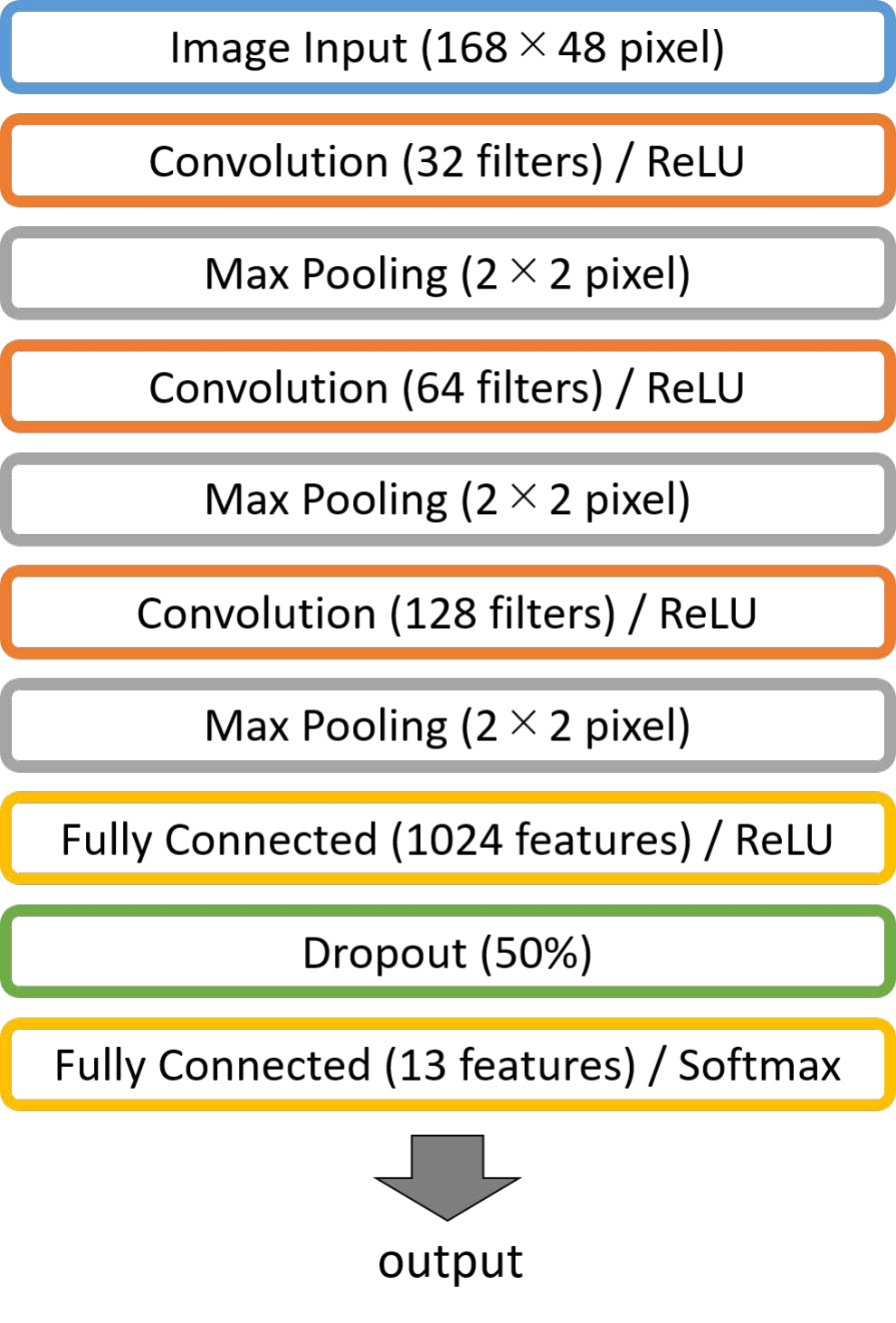}
  \caption{ Convolutional neural network (CNN) structure.}
  \label{tab:cnn}
\end{figure}

We used a CNN as a classifier for recognizing gestures, and the structure of that CNN is shown in Figure~\ref{tab:cnn}. For input, we used obtained 23 $\times$ 60 pixel images for the built-in sensor and 168 $\times$ 48 pixel images for the cover-type sensor for each gesture, as described in Section 4.2. The number of filters in each convolutional layer were 32, 64, and 128, and the filters were size 13 with a stride of 1. Batch normalization was performed in the convolutional layers to increase speed and stability of learning. Rectified linear units (ReLU) was the activation function used. The max pooling layers had a kernel size of 2 and stride of 2. The final output values were the probabilities that the gesture was classified into each of the classes. The CNN was implemented using TFLearn.\footnote{http://tflearn.org/}.

%% file: Sections/Evaluation.tex
\section{Evaluation}
We conducted two evaluation studies to investigate the recognition accuracy for the set of cushion gestures designed in the gesture elicitation study (Figure~\ref{tab:gesture}). 
For those studies, the sensor devices were connected by cable to a PC with the following specifications, and both studies used the same PC.

\begin{itemize}
    \item CPU: Intel(R) Core(TM) i7-8650U CPU @ 1.90 GHz 2.11 GHz
    \item OS: Windows 10 Home
    \item Python version: 3.6
\end{itemize}

\subsection{Evaluation \#1: Build-In Sensor}

There were 10 participants (7 males and 3 females) aged 21 to 25 years (average age, 23) for this experiment, where participants were asked to perform the 13 gestures using a cushion. The gestures were performed in random order, and a researcher instructed participants on when and which gestures to perform. All participants were right-handed. The cushion used was the same as that used in the experiments selecting the gestures.

Each gesture was described to participants, and they were allowed to practice the gestures on the cushion for approximately 2 minutes before the gestures were recorded. Each gesture was performed 10 times, so participants performed a total of 130 gestures. Participants were asked to place the cushion on their lap while recording each gesture. The experiment yielded 1,300 data points from 10 trials, 13 gestures, and 10 participants, and results were recorded for the two cases. For the first case, the data for each user were partitioned into the first half and the second half and used for, respectively, training and testing. For the second case, training was performed with all but one of the 10 users, and the excluded user was used for testing. In the first case, we presumed there would be differences in the gestures for each user, so we wanted to study whether learning models would need to be tuned for individuals. In the second case, we wanted to study the accuracy of a model trained with data from users other than the current user, examining the issue of the time required to train individualized models, as in the first case.

\subsubsection{Results \#1}
\begin{figure}[htbp]
  \begin{tabular}{cc}

\begin{minipage}{0.5\linewidth}
  \centering
  \includegraphics[width=\linewidth]{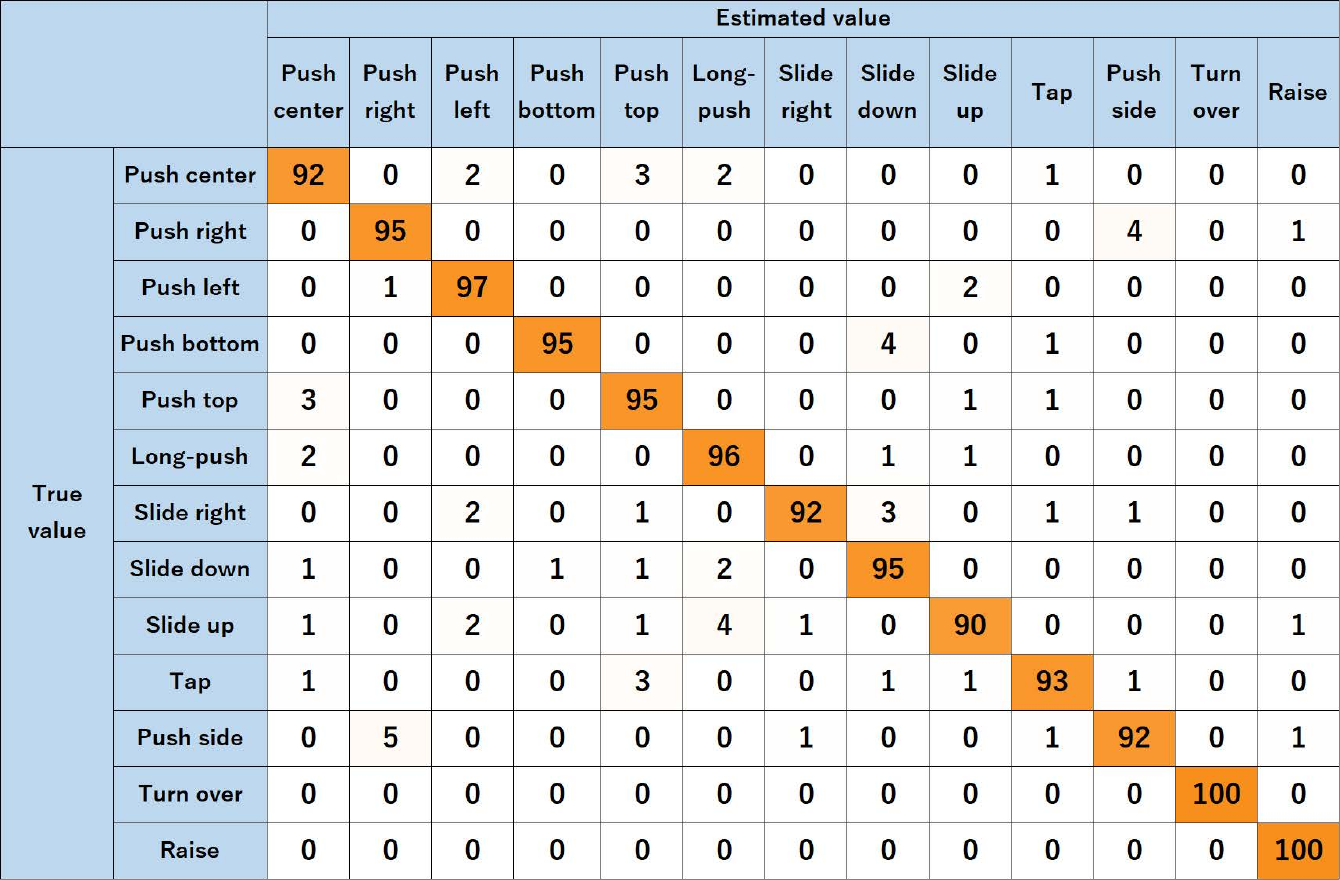}
  \caption{Confusion Matrix (per-subject training)}
  \label{tab:table3}
\end{minipage}

\begin{minipage}{0.5\linewidth}
  \centering
  \includegraphics[width=\linewidth]{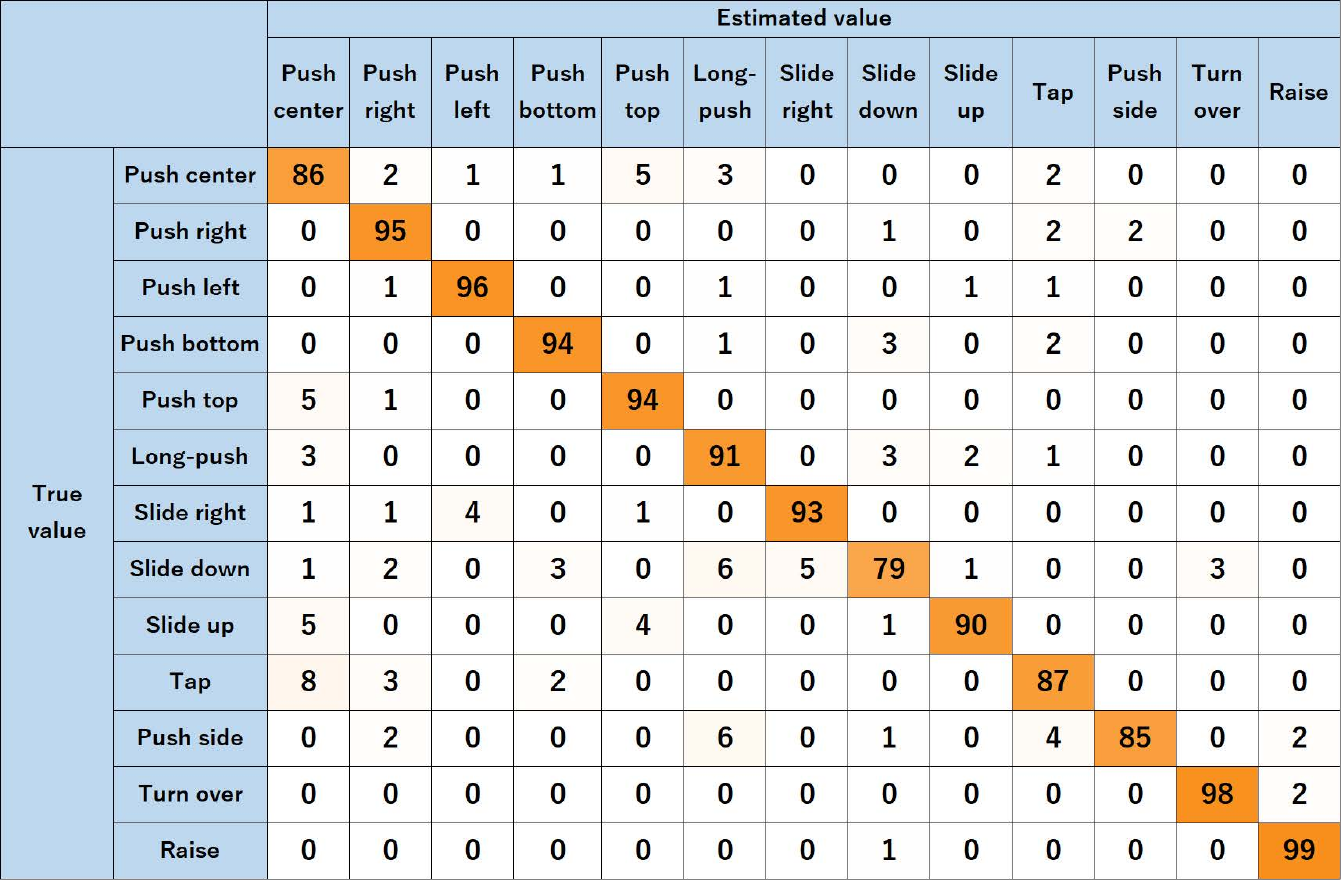}
  \caption{Confusion Matrix (raining with data from 9 subjects)}
  \label{tab:table4}
\end{minipage}

  \end{tabular}
\end{figure}

The recognition rates during training and testing for each user are shown in Figure~\ref{tab:table3}. Twofold cross-validation was used for the evaluation. the results indicated an average recognition accuracy of 94.8\%. There was some confusion between the ``push right'' and ``push right side'' gestures. This may have been partly because for both of these gestures, the density of the fibers near the right-of-center sensor module in the cushion increases. Gestures with the highest recognition accuracy were ``turn over'' and ``extend both arms and raise.'' All other gestures were done on the user’s lap, but these two gestures involve holding up the cushion and changing its orientation, so values from the accelerometers change greatly. This creates strong features in the images obtained, which may account for the accurate recognition.

The recognition rates when training with data from all but one user, and then testing with the data from the excluded user (leave-one-out cross-validation) are shown in Figure~\ref{tab:table4}. The average recognition rate was 91.3\%, and 8.0\% of ``tap with both hands'' cases were classified as ``push center.'' During the experiment, we noticed that the ``tap with both hands'' gesture varied in strength and length by participant. Although participants were instructed regarding the gestures before the experiment, it seems that it was relatively difficult for users to perform this gesture with consistent strength and length. When the strength or length of taps was stronger or longer than other users, it is possible that the data appeared similar to pushing the center of the cushion, increasing the likelihood that it would be recognized as such. To differentiate these gestures, it will be necessary to increase the number of sensor modules, improving the spatial ability to comprehend changes in the shape of the cushion. If there is still difficulty recognizing the gestures, it may necessary to assign different gestures.

In Figure~\ref{tab:table4}, the recognition rate for ``slide down'' is relatively low among the gestures. A possible reason for this drop in accuracy, which we did not initially anticipate, is that for slide gestures, different people apply pressure with different areas of their palms. If they apply pressure with the heel of the palm, it can be more difficult to apply pressure for the ``slide down'' gesture than for the ``slide up'' gesture, and this could have resulted in confusion with other gestures. 

\subsection{Evaluation \#2: Cover-Type Sensor}

There were 3 male and 2 female participants ranging in age from 16 to 47 (average age, 26) for this experiment. We used the user-defined gesture set that was designed to control home appliances \cite{Alexander}, and participants were given instructions on when and which gesture to perform. All participants were right-handed. Each gesture was performed 20 times, so each participant performed a total of 260 gestures. Participants were asked to place the cushion on the table while recording each gesture. The experiment yielded 1,300 data points from 20 trials, 13 gestures, and 5 participants.

\subsubsection{Results \#2}
\begin{figure}[htbp]
  \begin{tabular}{cc}

\begin{minipage}[t]{0.5\linewidth}
  \centering
  \includegraphics[width=\linewidth]{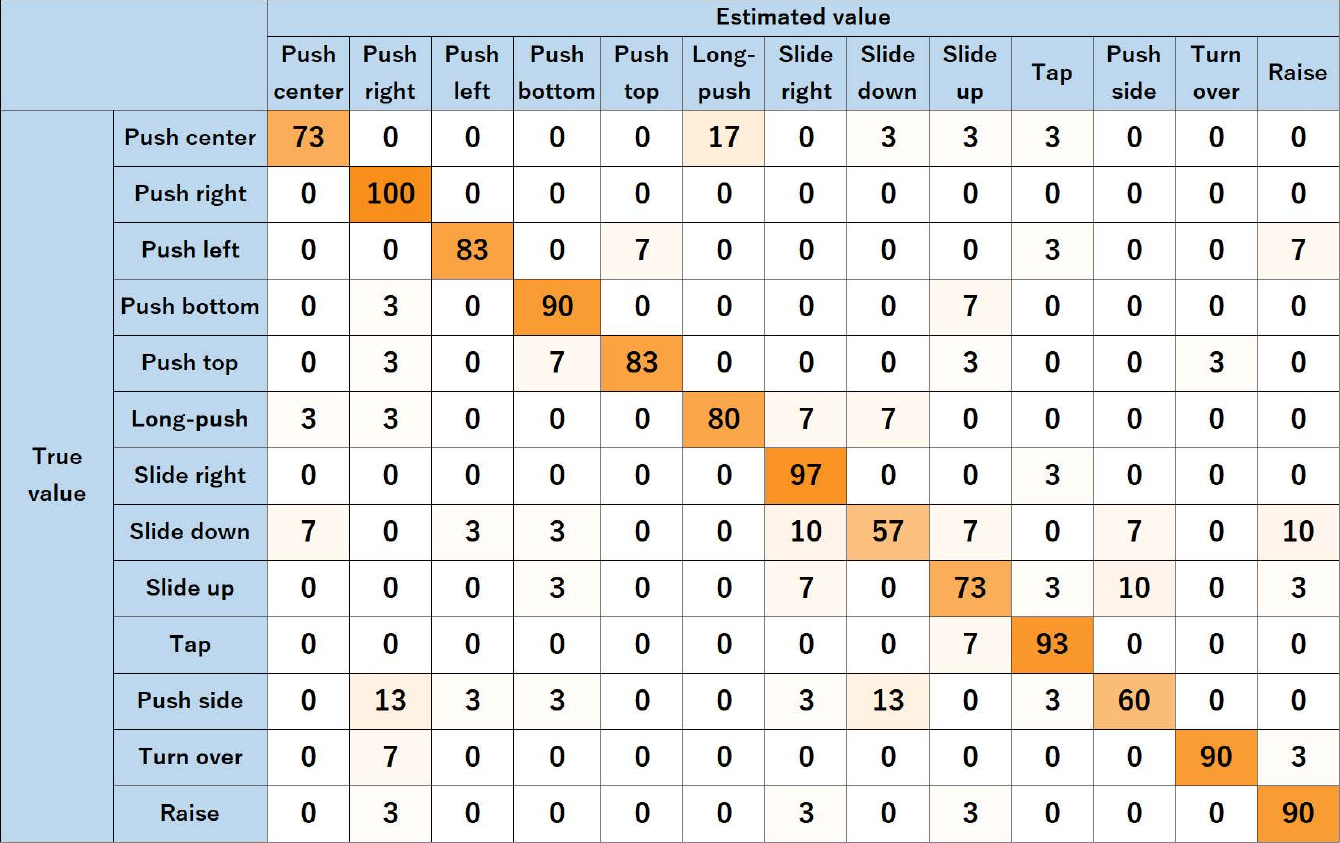}
  \caption{Confusion Matrix (per-subject training)}
  \label{tab:table5}
\end{minipage}

\begin{minipage}[t]{0.5\linewidth}
  \centering
  \includegraphics[width=\linewidth]{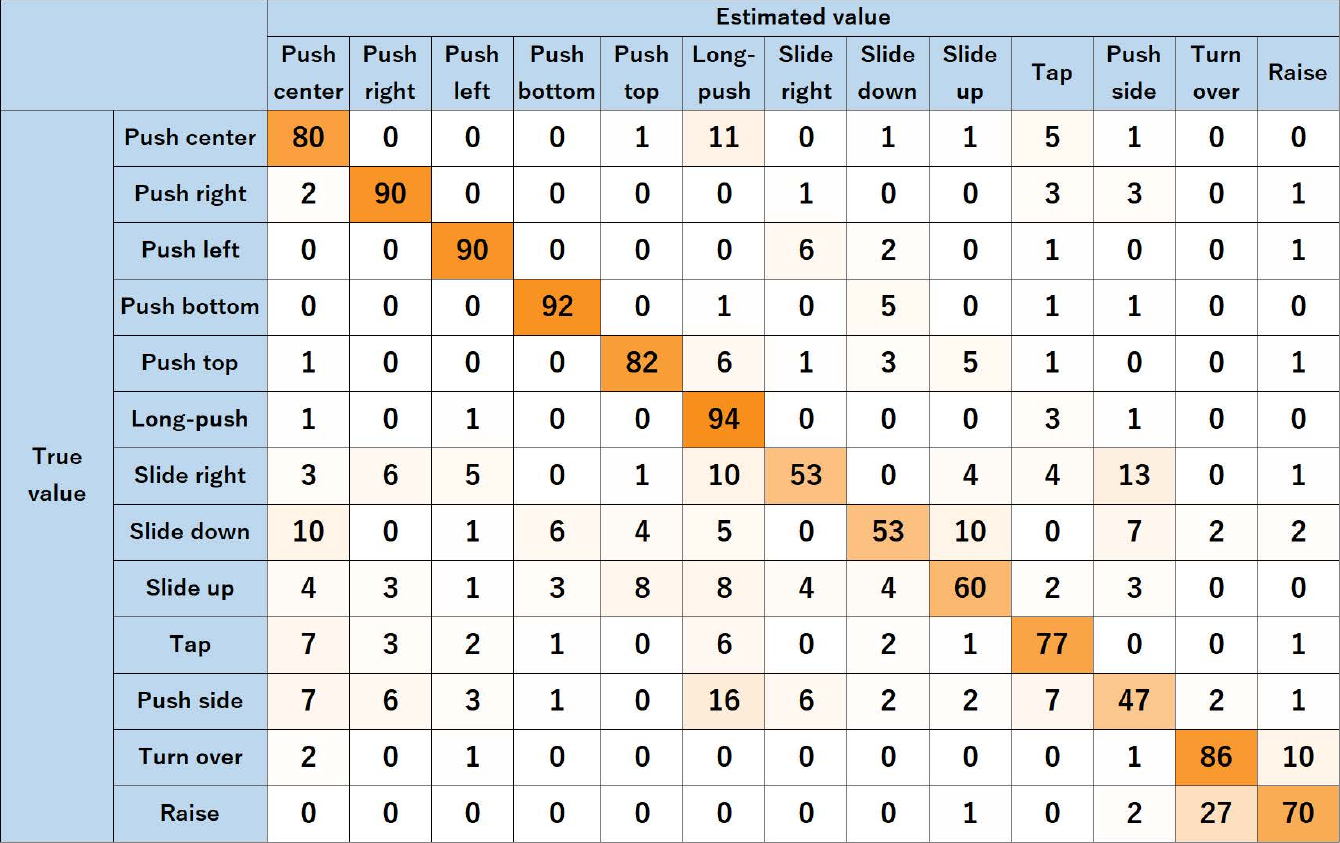}
  \caption{Confusion Matrix (training with data from 4 subjects)}
  \label{tab:table6}
\end{minipage}

  \end{tabular}
\end{figure}

The recognition rates for gestures when training and testing for each user are shown in Figure~\ref{tab:table5}. The results indicated an average recognition accuracy rate of 82.3\%. There was some confusion between the ``push center'' and ``long push center'' gestures. This is because the long push time was different for each person, and some participants did not show consistent lengths of time pushing the center. The ``slide down'' and ``push right side'' gestures showed less than 70\% accuracy. For the ``slide down'' gesture, we consider the reason for low accuracy to be that the area of force application, such as the front side or back side of the cushion, was not fixed. We think the same explanation is true for the ``slide up'' gesture. For the ``push right side'' gesture, we suspect small fluctuations in sensor values are the reason for low accuracy. On the other hand, the push right/left/up/bottom, tap, and move the cushion gestures showed high accuracy. Excluding gestures with less than 73\% accuracy, the average recognition accuracy of nine gestures was 89.6\%. 

The recognition rates when training with data from all but one user, and then testing with the data from the excluded user (leave-one-out cross-validation), are shown in Figure~\ref{tab:table6}. The average recognition rate was 74.9\%. There was some confusion between the ``turn over'' and ``extend both arms and raise'' gestures, which require similar movements toward the middle. For those who gesture slowly, we consider that the gesture does not fit within the set frame, and therefore, the sensor value does not get to the end. All ``slide'' gestures showed 60\% accuracy or less. We consider that these gestures tend to be executed at different strengths and speeds depending on the person. We think the same explanation is true for the ``push right side'' gesture.

%% file: Sections/Discussion.tex
\section{Discussion}


From the two evaluation studies, we found that the gesture-recognition accuracy of the built-in sensor was better than the cover-type sensor. While maintaining the comfort and functionality of the cushions, we need to consider the number and type of sensors. The advantage of the cover-type sensor is that once a user owns the interface, any cushion can be easily converted.

The average accuracy of 94.8\% with the built-in sensor was accurate enough for practical usage, which is also true of the 91.3\% result from the leave-one-out cross-validation. while there was one gesture with a recognition rate lower than 80\%, the overall recognition accuracy for the built-in sensor was high. Conversely, the average accuracy of the cover-type sensor was 82.3\%, which was not adequate for practical use. There were some gestures with 90+\% (including 100\%) accuracy, but there was also a case of lower than 60\%. The average cross-validation was also low (74.9\%). We tried to use the cover-type sensor for the purposes of convenience already discussed, but there was a trade-off between the accuracy and satisfaction of use. We may use the built-in sensor for the accurate control, and may use the cover-type sensor for the enjoyment.

We also observed that both sensors were stable with recognition for gestures involving pushing one of the four sides. However, sliding and stroking gestures on the surface were not accurately recognized. Those gestures on the rigid surfaces were stably recognized in previous studies~\cite{10.1145/1449715.1449747}, so we found that those gestures were weak in the case of sliding and striking fingers and hands on the surface of the soft object, especially measuring the changes to the surface with the acceleration sensor. The built-in sensors worked well to measure them but were also slightly weak in comparison to the pushing gestures.
%
%



\subsection{Possible Application}
We think that the cushion interface is easy to use for people with physical disabilities. People who are blind or unable to move their hands and fingers as much likely find it difficult to use an existing remote controller because of the fine control requirements. However, with a cushion interface, there is no need for detailed manipulation, and they can use actions such as pushing that they use on a daily basis. Moreover, it is soft and safe, even if a patient with cerebral palsy unintentionally cushions a part of his or her body too tightly due to involuntary movements. This interface can be safely used by anyone, not just people with disabilities. For example, children could use it as a game controller, and the elderly could use it as an exercise tool for their health.

\subsection{Design Issues}
One issue with the proposed built-in sensor method is the size of the modules. In these experiments, we used modules that are 4.4 cm $\times$ 6.8 cm $\times$ 3.3 cm, which is relatively small compared to the 43 cm $\times$ 43 cm cushions. However, if these modules were used in a smaller cushion, the distances between reflective optical sensors would be smaller, which could cause them to affect each other. Users could also find that the cushions feel stiff. The size of modules could be reduced by using a smaller microcontroller or by developing dedicated components.

Hysteresis could also become an issue when using a cushion for long periods of time. All cushions will change shape when pushed or tapped by users over long periods of time, and there is a chance that internal density distribution, elasticity, and other properties could change. In the future, further experiments will be needed to clarify reproducibility when a system is used over a long period.

There are also several issues that will need to be resolved with the technology in this study before implementing it in general use. In this study, we defined a gesture for switching the device being operated, but we did not design any feedback to indicate to the user which device is currently being controlled. Thus, some form of audible or visual feedback will be necessary in the future. There are several conceivable forms that such feedback could take, such as having the appliance provide the feedback, having an integrated terminal such as a television provide the feedback, or having the feedback built into the cushion. In any case, some form of feedback for switching the item being controlled will be needed. Such functionality is an issue for future consideration.

\subsection{Limitations}
Our work is a proof-of-concept study, so there are various limitations for practical usage. 
First, the batteries are a limitation in terms of how to charge the module batteries. There may be ways to mitigate this issue, such as conserving power by only turning on the infrared light when the cushion is in use, but some method of charging the batteries will be needed. Some form of operation, such as using wireless charging or self-charging by shaking the cushion, will need to be devised in the future as well, and this is another issue for consideration.

We used a 43 cm $\times$ 43 cm cushion in this study. Many cushion products in sizes ranging from 40 to 50 cm can be purchased at retail outlets in Japan, and the knowledge gained in this research can be applied to cushions in this range. However, with cushions outside of this range, it may be necessary to study changes in the size of the sensor modules or in the learning methods. We will continue to consider issues of size in the future.

There are also issues with the directionality of the cushion. Currently, our experiments yielded the stated performance only when gestures were performed with the cushion in a specific orientation. However, in everyday environments, they can be expected to be placed in many different orientations. Some possible ways to solve this issue for future consideration include a visual indication of the orientation on the cushion’s surface, entering the orientation before performing any operations, or collecting gesture data for multiple orientations.


%% file: Sections/Conclusion.tex
\section{Conclusion}

In this study, we developed a system that uses cushions, which are common in our surroundings, as an interface for controlling home appliances. For our method, we first conducted a gesture elicitation study to collect gestures suitable for controlling home appliances with a cushion, and obtained 13 gestures. We then proposed a method for recognizing time sequence gestures, which uses sensor modules consisting of reflective optical sensors and accelerometers along with a CNN. We conducted experiments to evaluate the accuracy of the gesture-recognition system, and they resulted in an average accuracy of 94.8\% when training for each user, and an average accuracy of 91.3\% when testing with a user that did not exist in the training data set. 


%% file: main.bbl

\begin{thebibliography}{26}


\ifx \showCODEN    \undefined \def \showCODEN     #1{\unskip}     \fi
\ifx \showDOI      \undefined \def \showDOI       #1{#1}\fi
\ifx \showISBNx    \undefined \def \showISBNx     #1{\unskip}     \fi
\ifx \showISBNxiii \undefined \def \showISBNxiii  #1{\unskip}     \fi
\ifx \showISSN     \undefined \def \showISSN      #1{\unskip}     \fi
\ifx \showLCCN     \undefined \def \showLCCN      #1{\unskip}     \fi
\ifx \shownote     \undefined \def \shownote      #1{#1}          \fi
\ifx \showarticletitle \undefined \def \showarticletitle #1{#1}   \fi
\ifx \showURL      \undefined \def \showURL       {\relax}        \fi
\providecommand\bibfield[2]{#2}
\providecommand\bibinfo[2]{#2}
\providecommand\natexlab[1]{#1}
\providecommand\showeprint[2][]{arXiv:#2}

\bibitem[\protect\citeauthoryear{Abeele, Zaman, and Abeele}{Abeele et~al\mbox{.}}{2008}]%
        {Aabeele}
\bibfield{author}{\bibinfo{person}{Vero~Vanden Abeele}, \bibinfo{person}{Bieke Zaman}, {and} \bibinfo{person}{Mariek~Vanden Abeele}.} \bibinfo{year}{2008}\natexlab{}.
\newblock \showarticletitle{The unlikeability of a cuddly toy interface: An experimental study of preschoolers’ likeability and usability of a 3D game played with a cuddly toy versus a keyboard}. In \bibinfo{booktitle}{\emph{International Conference on Fun and Games}}. \bibinfo{pages}{118--131}.
\newblock


\bibitem[\protect\citeauthoryear{Alexander, Han, Judd, Irani, and Subramanian}{Alexander et~al\mbox{.}}{2012}]%
        {Alexander}
\bibfield{author}{\bibinfo{person}{Jason Alexander}, \bibinfo{person}{Teng Han}, \bibinfo{person}{William Judd}, \bibinfo{person}{Pourang Irani}, {and} \bibinfo{person}{Sriram Subramanian}.} \bibinfo{year}{2012}\natexlab{}.
\newblock \showarticletitle{Putting your best foot forward: Investigating real-world mappings for foot-based gestures}. In \bibinfo{booktitle}{\emph{Proceedings of the SIGCHI Conference on Human Factors in Computing Systems}} \emph{(\bibinfo{series}{CHI '12})}. \bibinfo{pages}{1229–1238}.
\newblock


\bibitem[\protect\citeauthoryear{Brumitt, Meyers, Krumm, Kern, and Shafer}{Brumitt et~al\mbox{.}}{2000}]%
        {Brumitt}
\bibfield{author}{\bibinfo{person}{Barry Brumitt}, \bibinfo{person}{Brian Meyers}, \bibinfo{person}{John Krumm}, \bibinfo{person}{Amanda Kern}, {and} \bibinfo{person}{Steven~A. Shafer}.} \bibinfo{year}{2000}\natexlab{}.
\newblock \showarticletitle{EasyLiving: Technologies for intelligent environments}. In \bibinfo{booktitle}{\emph{Proceedings of the 2nd International Symposium on Handheld and Ubiquitous Computing (HUC)}}.
\newblock


\bibitem[\protect\citeauthoryear{Fukahori, Sakamoto, and Igarashi}{Fukahori et~al\mbox{.}}{2015}]%
        {Fukahori}
\bibfield{author}{\bibinfo{person}{Koumei Fukahori}, \bibinfo{person}{Daisuke Sakamoto}, {and} \bibinfo{person}{Takeo Igarashi}.} \bibinfo{year}{2015}\natexlab{}.
\newblock \showarticletitle{Exploring subtle foot plantar-based gestures with sock-placed pressure sensors}. In \bibinfo{booktitle}{\emph{Proceedings of the 33rd Annual ACM Conference on Human Factors in Computing Systems}}. \bibinfo{pages}{3019--3028}.
\newblock


\bibitem[\protect\citeauthoryear{Fukui, Okishiba, Karasawa, and Warisawa}{Fukui et~al\mbox{.}}{2017}]%
        {Fukui}
\bibfield{author}{\bibinfo{person}{Rui Fukui}, \bibinfo{person}{Shunsuke Okishiba}, \bibinfo{person}{Hiroyuki Karasawa}, {and} \bibinfo{person}{Shinichi Warisawa}.} \bibinfo{year}{2017}\natexlab{}.
\newblock \showarticletitle{Dynamic hand motion recognition based on wrist contour measurement for a W wearable display}. In \bibinfo{booktitle}{\emph{The Proceedings of JSME Annual Conference on Robotics and Mechatronics (Robomec)}}. \bibinfo{pages}{2A2--L02}.
\newblock


\bibitem[\protect\citeauthoryear{Harrison and Hudson}{Harrison and Hudson}{2008}]%
        {10.1145/1449715.1449747}
\bibfield{author}{\bibinfo{person}{Chris Harrison} {and} \bibinfo{person}{Scott~E. Hudson}.} \bibinfo{year}{2008}\natexlab{}.
\newblock \showarticletitle{Scratch input: Creating large, inexpensive, unpowered and mobile finger input surfaces}. In \bibinfo{booktitle}{\emph{Proceedings of the 21st Annual ACM Symposium on User Interface Software and Technology}} \emph{(\bibinfo{series}{UIST '08})}. \bibinfo{publisher}{ACM}, \bibinfo{address}{New York, NY}, \bibinfo{pages}{205–208}.
\newblock
\showISBNx{9781595939753}
\urldef\tempurl%
\url{https://doi.org/10.1145/1449715.1449747}
\showDOI{\tempurl}


\bibitem[\protect\citeauthoryear{Ikeda, Koizumi, and Naemura}{Ikeda et~al\mbox{.}}{2017}]%
        {Ikeda}
\bibfield{author}{\bibinfo{person}{Kohei Ikeda}, \bibinfo{person}{Naoya Koizumi}, {and} \bibinfo{person}{Takeshi Naemura}.} \bibinfo{year}{2017}\natexlab{}.
\newblock \showarticletitle{FunCushion: Fabricating functional cushion interfaces with fluorescent-pattern displays}. In \bibinfo{booktitle}{\emph{International Conference on Advances in Computer Entertainment}}. \bibinfo{pages}{470--487}.
\newblock


\bibitem[\protect\citeauthoryear{Irie, Wakamura, and Umeda}{Irie et~al\mbox{.}}{2007}]%
        {Irie}
\bibfield{author}{\bibinfo{person}{Kota Irie}, \bibinfo{person}{Naohiro Wakamura}, {and} \bibinfo{person}{Kazumori Umeda}.} \bibinfo{year}{2007}\natexlab{}.
\newblock \showarticletitle{Construction of an intelligent room based on gesture recognition}.
\newblock \bibinfo{journal}{\emph{Transactions of the Japan Society of Mechanical Mngineers. C}} \bibinfo{volume}{73}, \bibinfo{number}{725} (\bibinfo{year}{2007}), \bibinfo{pages}{258--265}.
\newblock


\bibitem[\protect\citeauthoryear{Johnson, Wilson, Blumberg, Kline, and Bobick}{Johnson et~al\mbox{.}}{1999}]%
        {Johnson}
\bibfield{author}{\bibinfo{person}{Michael~Patrick Johnson}, \bibinfo{person}{Andrew Wilson}, \bibinfo{person}{Bruce Blumberg}, \bibinfo{person}{Christopher Kline}, {and} \bibinfo{person}{Aaron Bobick}.} \bibinfo{year}{1999}\natexlab{}.
\newblock \showarticletitle{Sympathetic interfaces: Using a plush toy to direct aynthetic characters}. In \bibinfo{booktitle}{\emph{Proceedings of the SIGCHI Conference on Human Factors in Computing Systems}} \emph{(\bibinfo{series}{CHI '99})}. \bibinfo{pages}{152–158}.
\newblock


\bibitem[\protect\citeauthoryear{Kidd, Orr, Abowd, Atkeson, Essa, MacIntyre, Mynatt, Starner, and Newstetter}{Kidd et~al\mbox{.}}{1999}]%
        {Kidd}
\bibfield{author}{\bibinfo{person}{Cory~D. Kidd}, \bibinfo{person}{Robert Orr}, \bibinfo{person}{Gregory~D. Abowd}, \bibinfo{person}{Christopher~G. Atkeson}, \bibinfo{person}{Irfan~A. Essa}, \bibinfo{person}{Blair MacIntyre}, \bibinfo{person}{Elizabeth~D. Mynatt}, \bibinfo{person}{Thad~E. Starner}, {and} \bibinfo{person}{Wendy Newstetter}.} \bibinfo{year}{1999}\natexlab{}.
\newblock \showarticletitle{The aware home: A living laboratory for ubiquitous computing research}. In \bibinfo{booktitle}{\emph{Proceedings of the Second International Workshop on Cooperative Buildings, Integrating Information, Organization, and Architecture (CoBuild)}}. \bibinfo{pages}{191--198}.
\newblock


\bibitem[\protect\citeauthoryear{Kikui, Itoh, Yamada, Sugiura, and Sugimoto}{Kikui et~al\mbox{.}}{2018}]%
        {Kikui}
\bibfield{author}{\bibinfo{person}{Kosuke Kikui}, \bibinfo{person}{Yuta Itoh}, \bibinfo{person}{Makoto Yamada}, \bibinfo{person}{Yuta Sugiura}, {and} \bibinfo{person}{Maki Sugimoto}.} \bibinfo{year}{2018}\natexlab{}.
\newblock \showarticletitle{Intra-/inter-user adaptation framework for wearable gesture sensing device}. In \bibinfo{booktitle}{\emph{Proceedings of the 2018 ACM International Symposium on Wearable Computers}}. \bibinfo{pages}{21--24}.
\newblock


\bibitem[\protect\citeauthoryear{K{\"u}hnel, Westermann, Hemmert, Kratz, M{\"u}ller, and M{\"o}ller}{K{\"u}hnel et~al\mbox{.}}{2011}]%
        {Kuhnel}
\bibfield{author}{\bibinfo{person}{Christine K{\"u}hnel}, \bibinfo{person}{Tilo Westermann}, \bibinfo{person}{Fabian Hemmert}, \bibinfo{person}{Sven Kratz}, \bibinfo{person}{Alexander M{\"u}ller}, {and} \bibinfo{person}{Sebastian M{\"o}ller}.} \bibinfo{year}{2011}\natexlab{}.
\newblock \showarticletitle{I'm home: Defining and evaluating a gesture set for smart-home control}.
\newblock \bibinfo{journal}{\emph{International Journal of Human-Computer Studies}} \bibinfo{volume}{69}, \bibinfo{number}{11} (\bibinfo{year}{2011}), \bibinfo{pages}{693--704}.
\newblock


\bibitem[\protect\citeauthoryear{Masui, Tsukada, and Siio}{Masui et~al\mbox{.}}{2004}]%
        {Masui}
\bibfield{author}{\bibinfo{person}{Toshiyuki Masui}, \bibinfo{person}{Koji Tsukada}, {and} \bibinfo{person}{Itiro Siio}.} \bibinfo{year}{2004}\natexlab{}.
\newblock \showarticletitle{MouseField: A Simple and Versatile Input Device for Ubiquitous Computing}. In \bibinfo{booktitle}{\emph{UbiComp2004: Ubiquitous Computing}}. \bibinfo{pages}{319--328}.
\newblock


\bibitem[\protect\citeauthoryear{Paiva, Chaves, Piedade, Bullock, Andersson, and H\"{o}\"{o}k}{Paiva et~al\mbox{.}}{2003}]%
        {Paiva}
\bibfield{author}{\bibinfo{person}{Ana Paiva}, \bibinfo{person}{Ricardo Chaves}, \bibinfo{person}{Mois\'{e}s Piedade}, \bibinfo{person}{Adrian Bullock}, \bibinfo{person}{Gerd Andersson}, {and} \bibinfo{person}{Kristina H\"{o}\"{o}k}.} \bibinfo{year}{2003}\natexlab{}.
\newblock \showarticletitle{SenToy: A tangible interface to control the emotions of a synthetic character}. In \bibinfo{booktitle}{\emph{Proceedings of the Second International Joint Conference on Autonomous Agents and Multiagent Systems}} \emph{(\bibinfo{series}{AAMAS '03})}. \bibinfo{pages}{1088–1089}.
\newblock
\showISBNx{1581136838}


\bibitem[\protect\citeauthoryear{Pomboza-Junez and A~Holgado-Terriza}{Pomboza-Junez and A~Holgado-Terriza}{2015}]%
        {Gonzalo}
\bibfield{author}{\bibinfo{person}{Gonzalo Pomboza-Junez} {and} \bibinfo{person}{Juan A~Holgado-Terriza}.} \bibinfo{year}{2015}\natexlab{}.
\newblock \showarticletitle{Control of home devices based on hand gestures}. In \bibinfo{booktitle}{\emph{2015 IEEE 5th International Conference on Consumer Electronics (ICCE-Berlin)}}. \bibinfo{pages}{510--514}.
\newblock


\bibitem[\protect\citeauthoryear{Seifried, Haller, Scott, Perteneder, Rendl, Sakamoto, and Inami}{Seifried et~al\mbox{.}}{2009}]%
        {10.1145/1731903.1731911}
\bibfield{author}{\bibinfo{person}{Thomas Seifried}, \bibinfo{person}{Michael Haller}, \bibinfo{person}{Stacey~D. Scott}, \bibinfo{person}{Florian Perteneder}, \bibinfo{person}{Christian Rendl}, \bibinfo{person}{Daisuke Sakamoto}, {and} \bibinfo{person}{Masahiko Inami}.} \bibinfo{year}{2009}\natexlab{}.
\newblock \showarticletitle{CRISTAL: A collaborative home media and device controller based on a multi-touch display}. In \bibinfo{booktitle}{\emph{Proceedings of the ACM International Conference on Interactive Tabletops and Surfaces}} (Banff, Alberta, Canada) \emph{(\bibinfo{series}{ITS '09})}. \bibinfo{publisher}{ACM}, \bibinfo{address}{New York, NY}, \bibinfo{pages}{33–40}.
\newblock
\showISBNx{9781605587332}
\urldef\tempurl%
\url{https://doi.org/10.1145/1731903.1731911}
\showDOI{\tempurl}


\bibitem[\protect\citeauthoryear{Seki, Sugiyama, Sudo, Nakano, and Hada}{Seki et~al\mbox{.}}{2014}]%
        {Seki}
\bibfield{author}{\bibinfo{person}{Megumi Seki}, \bibinfo{person}{Nozomi Sugiyama}, \bibinfo{person}{Atsuhito Sudo}, \bibinfo{person}{Akito Nakano}, {and} \bibinfo{person}{Hisakazu Hada}.} \bibinfo{year}{2014}\natexlab{}.
\newblock \showarticletitle{A proposal of the stuffed toy type interface for smart houses}. In \bibinfo{booktitle}{\emph{Proceedings of Entertainment Computing 2014}}.
\newblock


\bibitem[\protect\citeauthoryear{Siio}{Siio}{2010}]%
        {Shiio}
\bibfield{author}{\bibinfo{person}{Itiro Siio}.} \bibinfo{year}{2010}\natexlab{}.
\newblock \showarticletitle{Recent Trends in Real-world Interface Systems : Ubiquitous Computing for Residence}.
\newblock \bibinfo{journal}{\emph{Journal of the Neurological Sciences}} \bibinfo{volume}{51}, \bibinfo{number}{7} (\bibinfo{year}{2010}), \bibinfo{pages}{795 -- 802}.
\newblock


\bibitem[\protect\citeauthoryear{Starner, Auxier, Ashbrook, and Gandy}{Starner et~al\mbox{.}}{2000}]%
        {Starner}
\bibfield{author}{\bibinfo{person}{Thad Starner}, \bibinfo{person}{Jake Auxier}, \bibinfo{person}{Daniel Ashbrook}, {and} \bibinfo{person}{Maribeth Gandy}.} \bibinfo{year}{2000}\natexlab{}.
\newblock \showarticletitle{The gesture pendant: A self-illuminating, wearable, infrared computer vision system for home automation control and medical monitoring}. In \bibinfo{booktitle}{\emph{Digest of Papers. Fourth International Symposium on Wearable Computers}}. \bibinfo{pages}{87--94}.
\newblock


\bibitem[\protect\citeauthoryear{Sugiura, Kakehi, Withana, Lee, Sakamoto, Sugimoto, Inami, and Igarashi}{Sugiura et~al\mbox{.}}{2011}]%
        {Sugiura}
\bibfield{author}{\bibinfo{person}{Yuta Sugiura}, \bibinfo{person}{Gota Kakehi}, \bibinfo{person}{Anusha Withana}, \bibinfo{person}{Calista Lee}, \bibinfo{person}{Daisuke Sakamoto}, \bibinfo{person}{Maki Sugimoto}, \bibinfo{person}{Masahiko Inami}, {and} \bibinfo{person}{Takeo Igarashi}.} \bibinfo{year}{2011}\natexlab{}.
\newblock \showarticletitle{Detecting Shape Deformation of Soft Objects Using Directional Photoreflectivity Measurement}. In \bibinfo{booktitle}{\emph{Proceedings of the 24th annual ACM symposium on User interface software and technology (UIST)}}.
\newblock


\bibitem[\protect\citeauthoryear{Tsukada and Yasumura}{Tsukada and Yasumura}{2004}]%
        {Tsukada}
\bibfield{author}{\bibinfo{person}{Koji Tsukada} {and} \bibinfo{person}{Michiaki Yasumura}.} \bibinfo{year}{2004}\natexlab{}.
\newblock \showarticletitle{Ubi-Finger: A simple gesture input device for mobile and ubiquitous environment}.
\newblock \bibinfo{journal}{\emph{Asian Information Science Life (AISL)}}  \bibinfo{volume}{2} (\bibinfo{year}{2004}), \bibinfo{pages}{111--120}.
\newblock


\bibitem[\protect\citeauthoryear{Ujima, Kadomura, and Siio}{Ujima et~al\mbox{.}}{2014}]%
        {Ujima}
\bibfield{author}{\bibinfo{person}{Kaori Ujima}, \bibinfo{person}{Azusa Kadomura}, {and} \bibinfo{person}{Itiro Siio}.} \bibinfo{year}{2014}\natexlab{}.
\newblock \showarticletitle{U-remo: Projection-assisted gesture control for home electronics}.
\newblock In \bibinfo{booktitle}{\emph{CHI'14 Extended Abstracts on Human Factors in Computing Systems}}. \bibinfo{pages}{1609--1614}.
\newblock


\bibitem[\protect\citeauthoryear{Vanderloock, Vanden~Abeele, Suykens, and Geurts}{Vanderloock et~al\mbox{.}}{2013}]%
        {Vanderloock}
\bibfield{author}{\bibinfo{person}{Karen Vanderloock}, \bibinfo{person}{Vero Vanden~Abeele}, \bibinfo{person}{Johan~AK Suykens}, {and} \bibinfo{person}{Luc Geurts}.} \bibinfo{year}{2013}\natexlab{}.
\newblock \showarticletitle{The skweezee system: enabling the design and the programming of squeeze interactions}. In \bibinfo{booktitle}{\emph{Proceedings of the 26th annual ACM symposium on User interface software and technology}}. \bibinfo{pages}{521--530}.
\newblock


\bibitem[\protect\citeauthoryear{Wobbrock, Morris, and Wilson}{Wobbrock et~al\mbox{.}}{2009}]%
        {Jacob}
\bibfield{author}{\bibinfo{person}{Jacob~O. Wobbrock}, \bibinfo{person}{Meredith~Ringel Morris}, {and} \bibinfo{person}{Andrew~D. Wilson}.} \bibinfo{year}{2009}\natexlab{}.
\newblock \showarticletitle{User-defined gestures for surface computing}. In \bibinfo{booktitle}{\emph{Proceedings of the SIGCHI Conference on Human Factors in Computing Systems (CHI)}}. \bibinfo{pages}{1083--1092}.
\newblock


\bibitem[\protect\citeauthoryear{Yagi, Kobayashi, Kashiwagi, Uriu, and Okude}{Yagi et~al\mbox{.}}{2011}]%
        {Yagi}
\bibfield{author}{\bibinfo{person}{Izumi Yagi}, \bibinfo{person}{Shigeru Kobayashi}, \bibinfo{person}{Ryo Kashiwagi}, \bibinfo{person}{Daisuke Uriu}, {and} \bibinfo{person}{Naohito Okude}.} \bibinfo{year}{2011}\natexlab{}.
\newblock \showarticletitle{Media cushion: Soft interface to control living environment using human natural behavior}.
\newblock In \bibinfo{booktitle}{\emph{ACM SIGGRAPH 2011 Posters}}. \bibinfo{pages}{1--1}.
\newblock


\bibitem[\protect\citeauthoryear{Yoshizawa, Yonezawa, and Iwai}{Yoshizawa et~al\mbox{.}}{2015}]%
        {Yoshizawa}
\bibfield{author}{\bibinfo{person}{Kazuhiro Yoshizawa}, \bibinfo{person}{Yoji Yonezawa}, {and} \bibinfo{person}{Masayuki Iwai}.} \bibinfo{year}{2015}\natexlab{}.
\newblock \showarticletitle{iRemocon-based Home Appliance Control Scheme using Gesture Information of Kinect}. In \bibinfo{booktitle}{\emph{Information Processing Society of Japan (IPSJ) Intaraction}}. \bibinfo{pages}{152--155}.
\newblock


\end{thebibliography}
